\documentclass[12pt] {article}
\usepackage{jheppub}
\usepackage{amsmath}
\usepackage{graphicx}
\usepackage{bm}
\usepackage{epsfig}
\usepackage[english]{babel}

\def\diag{\mbox{diag}}

\def\ab{\mathrm{abb.}}

\def\th{\mathrm{thr.}}

\def\MeV{~\mathrm{MeV}}
\def\GeV{~\mathrm{GeV}}
\title{Replaying neutrino bremsstrahlung with general dispersion relations}
\author[a,b,c]{Miao Li}
\author[a,b]{Da Liu}
\author[a,b]{Jun Meng}
\author[d,b]{Tower Wang}
\author[a,b]{Lanjun Zhou}
\affiliation[a]{Institute of Theoretical Physics, Chinese Academy of Sciences,\\
Beijing 100080, China\\}
\affiliation[b]{Kavli Institute for Theoretical Physics China, Chinese Academy of Sciences,\\
Beijing 100080, China\\}
\affiliation[c]{State Key Laboratory of Theoretical Physics, Institute of Theoretical Physics, Chinese Academy of Sciences,\\
Beijing 100190, China\\}
\affiliation[d]{Department of Physics, East China Normal University,\\
Shanghai 200241, China\\}
\emailAdd{mli@itp.ac.cn,liuda@itp.ac.cn,twang@phy.ecnu.edu.cn}
\abstract{It is generally held that neutrinos with superluminal
velocity will lose their energy spontaneously by radiating
electron-positron pairs, similar to bremsstrahlung process.
Recently, this process was closely studied for neutrinos whose
energy is roughly proportional to their momentum. Confronted with an
increasing amount of superluminal neutrino models, it is urgent to
calculate the same process for general dispersion relations. The
calculation is performed in this paper, without resorting to any
nontrivial frame such as the effective ``rest frame''.}
\keywords{Lorentz violation, neutrino, special relativity.}
\begin{document}
\maketitle
\section{Introduction}\label{sec:intr}

The OPERA experiment stirred the physics community recently with its
astonishing result that neutrinos in this experiment travel
apparently faster than light at a high confidence level
\cite{OPERA:2011zb}. If not attributed to systematic errors in the
measurement, this result would imply the violation of special
relativity. Subsequently, it has inspired a lot of
speculation.\footnote{As a partial list, see
\cite{Pfeifer:2011ve,Alexandre:2011bu,Wang:2011sz,Franklin:2011ws,Dass:2011yj,Winter:2011zf,Wang:2011zk,Li:2011zm,Aref'eva:2011zp,Saridakis:2011eq,Nojiri:2011ju,Zhu:2011fx,Li:2011rt,Qin:2011md,Bramante:2011uu,Zhao:2011sb,Chang:2011td,Matone:2011fn,Evslin:2011vq}
and references therein for speculation on this issue from various
aspects.}

However, as quickly claimed in \cite{Cohen:2011hx,Bi:2011nd},
several high-energy processes disfavor the superluminal
interpretation of the OPERA data. An outstanding example is the
bremsstrahlung-like process
\begin{equation}\label{bremsstrahlung}
\nu_{\mu}\rightarrow\nu_{\mu}+e^{+}+e^{-},
\end{equation}
where electron-positron pairs are radiated and hence neutrinos lose
their energy efficiently. To explicitly demonstrate this point, the
authors of \cite{Cohen:2011hx} assumed a special dispersion relation
roughly of the form $E=v_{\nu}p$, where $v_{\nu}$ is a constant
greater than light velocity. A similar assumption was also taken in
\cite{Bi:2011nd}.

On the other hand, a dispersion relation of the form $E=v_{\nu}p$ is
too oversimplified to accommodate more observational data of
neutrino velocity, as summarized and analyzed in \cite{Li:2011ue}.
Therefore, it is urgent to extend the calculation of
\cite{Cohen:2011hx,Bi:2011nd} to general dispersion relations. The
present paper is devoted to such a calculation. Our results would
help to rule out more phenomenological models and hunt for viable
models.

In the absence of Lorentz invariance, the calculation is complicated
even for the special dispersion relation $E=v_{\nu}p$, so an
effective ``mass'' was assigned to neutrinos and a ``rest frame''
was employed in \cite{Cohen:2011hx}. Our calculation does not resort
to such a nontrivial reference frame. Or more explicitly, one may
interpret our work as a direct calculation in the laboratory frame.
Applied to the above dispersion relation, our result provides a
crosscheck for the results in \cite{Cohen:2011hx,Bi:2011nd}.

The paper is organized as follows. Section \ref{sec:assum} collects
the basic assumptions and conventions of notation in this work.
Kinematically there is a threshold energy for the process
\eqref{bremsstrahlung}. We discuss the dependence of this threshold
on dispersion relations in section \ref{sec:thres}. Section
\ref{sec:width} is the main part of our paper, where we calculate
the ``decay width'' of superluminal neutrino via
\eqref{bremsstrahlung}. Details and techniques are presented
clearly. To check and apply our general results in sections
\ref{sec:thres} and \ref{sec:width}, we work out some specific
examples in section \ref{sec:exam} with given dispersion relations.
The results are consistent with \cite{Cohen:2011hx} quantitatively
and \cite{Mohanty:2011rm} qualitatively. Section \ref{sec:con}
concludes this paper.

\section{Basic assumptions and notation conventions}\label{sec:assum}

Throughout the paper, we assume
\begin{enumerate}
\item The ordinary conservation law for energy and momentum is
intact. In other words, the time and space translations are exact
symmetries in the working frame. A case study for violating this
assumption can be found in \cite{AmelinoCamelia:2011bz}.
\item The space is Euclidean and isotropic. Thereby we can work in a
spherical coordinate system and define the magnitude of momentum as
$p=|\vec{p}|$.
\item In the relevant energy range, the dispersion relation of electron
and positron is well characterized by $E^2=p^2+m_e^2$. This
assumption is in accordance with experiments to date.
\item The dispersion relation of neutrino
is either $E=E(p)$ or $p=p(E)$. Here $E(p)$ and $p(E)$ are arbitrary
devisable functions. They could be non-monotonic and may involve
some parameters such as mass, etc.
\end{enumerate}

In section \ref{sec:thres}, when deriving the threshold energy of
process \eqref{bremsstrahlung}, we make one more assumption:
\begin{itemize}
\item The neutrino's dispersion relation reduces to $E^2=p^2+m_{\nu}^2$
at very low energies, typically lower than the threshold energy by a
factor $\mathcal{O}(10^{-4})$. See details in section
\ref{sec:thres}.
\end{itemize}
In section \ref{sec:width}, this assumption is replaced by two other
assumptions:
\begin{itemize}
\item The squared amplitude \eqref{amplitude} is the same as that in
standard model. We will comment on possible loopholes of this
assumption in section \ref{sec:con}. But to alleviate the complexity
of our calculation, we should make such an assumption at the moment.
\item The masses of electron and positron are neglected. This assumption
seems to be reasonable, because when we study the decay width of
high-energy neutrino which is practically around or above $\GeV$
scale, the phase space should be dominated by high-energy particles
in principle.
\end{itemize}

Let us clarify the conventions of notation by writing
\eqref{bremsstrahlung} in the form
\begin{equation}
\nu(p)\rightarrow\nu(p')+e^{+}(k')+e^{-}(k),
\end{equation}
where we have specified the notation of momentum for each particle.
One must be careful of the notations of momenta. Taking $p$ for
instance, sometimes it stands for the four-vector, but sometimes it
stands for the magnitude of three-vector $\vec{p}$. In our
equations, the four-vectors appear usually together with a dot,
indicating the inner product with Lorentz signature
$\diag(+,-,-,-)$. For example, $p\cdot p=E_{p}^2-|\vec{p}|^2$ but
$p^2=|\vec{p}|^2$. So there are no confusions if one is careful.

The velocity of light is a constant, so we set it to $1$ throughout
this paper.

\section{Energy threshold}\label{sec:thres}

The assumptions made in the previous section simplify the derivation
of threshold energy for \eqref{bremsstrahlung}
considerably.\footnote{When our work was in preparation, ref.
\cite{Villante:2011pk} appeared. The topics in this section and
\cite{Villante:2011pk} overlap partly. For the completeness of our
paper, we keep this section in its own form. Ref.
\cite{Mohanty:2011rm}, starting form different assumptions, also
concerns a partly overlapped subject of this paper. It appeared more
recently when we were polishing our work.} From the energy
conservation relation
\begin{equation}\label{econs}
E_{p}=E_{p'}+E_{k'}+E_{k},
\end{equation}
we observe that to minimize $E_{p}$, one should lower $E_{p'}$,
$E_{k'}$ and $E_{k}$ as much as possible. This can be achieved by
deleting all of the transverse momentum components, leading to the
reduced conservation law of momentum
\begin{equation}\label{pcons}
p=p'+k'+k,
\end{equation}
where $p=|\vec{p}|$, $p=|\vec{p'}|$ and so on are magnitude of
momenta as our conventions.

Substituting \eqref{econs} and \eqref{pcons} into the dispersion
relation of $e^{+}(k')$ and $e^{-}(k)$, we have
\begin{equation}\label{Epfun}
(E_{p}-E_{p'})^2-2E_{k}(E_{p}-E_{p'})-(p-p')^2+2k(p-p')=0.
\end{equation}
In light of dispersion relations of $\nu(p)$, $\nu(p')$ and
$e^{-}(k)$, this equation may be understood as an implicit function
of $E_{p}$, $E_{p'}$ and $E_{k}$. Then we can extremize $E_{p}$ with
respect to $E_{p'}$ and $E_{k}$, obtaining
\begin{eqnarray}\label{extrem}
(E_{p}-E_{k})-(p-k)\frac{E_{p'}}{p'}&=&0,\nonumber\\
(E_{p}-E_{p'})-(p-p')\frac{E_{k}}{k}&=&0,
\end{eqnarray}
where conditions $\partial E_{p}/\partial E_{p'}=\partial
E_{p}/\partial E_{k}=0$ and dispersion relations
$dp'/dE_{p'}=E_{p'}/p'$, $dk/dE_{k}=E_{k}/k$ are used.

From eqs. \eqref{Epfun} and \eqref{extrem}, it is not hard to get
\begin{equation}
2E_{k}=E_{p}-E_{p'},\qquad2k=p-p',\qquad\frac{E_{p}}{p}=\frac{E_{p'}}{p'}
\end{equation}
and subsequently
\begin{equation}\label{midstep}
E_{p}^2-2E_{p}E_{p'}+m_{\nu}^2=p^2-2pp'+4m_e^2,\qquad E_{p'}=\frac{m_{\nu}}{\sqrt{1-\frac{p^2}{E_{p}^2}}}.
\end{equation}
As a result, we find the threshold of \eqref{bremsstrahlung} is
given by
\begin{equation}\label{threshold}
\left(E_{p}^2-p^2\right)_{\th}=\left(2m_{e}+m_{\nu}\right)^2.
\end{equation}
This result is in accordance with \cite{Villante:2011pk}. To the
leading order, it is also compatible with \cite{Cohen:2011hx} where
neutrino's dispersion relation is $E_{p}/p=v_{\nu}$. This will be
shown in subsection \ref{subsec:cg}. As a concrete example, in
subsection \ref{subsec:lw} we will utilize \eqref{threshold} to get
the decay threshold for a toy model of mass-dependent Lorentz
violation \cite{Li:2011ue}.

Note that when writing down \eqref{extrem} and \eqref{midstep}, we
have assumed the dispersion relation $E_{p'}^2=p'^2+m_{\nu}^2$ of
$\nu(p')$ at low energies. But the dispersion relation of $\nu(p)$
is irrelevant throughout the derivation. From eqs. \eqref{midstep}
and \eqref{threshold} we can see
$(E_{p'}/E_{p})_{\th}=m_{\nu}/(2m_{e}+m_{\nu})\sim\mathcal{O}(10^{-4})$.
That means we have assumed the dispersion relation
$E_{p'}^2=p'^2+m_{\nu}^2$ for neutrinos at an energy lower than the
threshold energy by $\mathcal{O}(10^{-4})$. This assumption is
natural and consistent with the observational data of neutrino
velocity summarized in \cite{Li:2011ue}. Replacing this assumption
with other dispersion relations, one may also restart from
\eqref{Epfun} and follow our method to derive the threshold
energy.\footnote{This method is valid if $E_{p}$ has local minima as
an implicit function \eqref{Epfun} of $E_{p'}$ and $E_{k}$. The
situation will be more complicated if the configuration of
\eqref{Epfun} does not have a local minimum.}

\section{Decay width}\label{sec:width}

We are interested in the following process:
$\nu(p)\rightarrow\nu(p')+e^{+}(k')+e^{-}(k)$. Considering the
neutral current of this process, we have\footnote{In the first
version of our manuscript, the squared amplitude is incomplete. Here
we corrected this error and included the last term in
\eqref{amplitude}. The difference does not affect most of our
calculations. It only modifies an overall factor in the decay width.
We are grateful to Zhaohuan Yu, Fedor Bezrukov and Evslin Jarah for
communications on this point.}
\begin{equation}
\sum_{spin}\mathcal{M}\mathcal{M}^{*}=128G_{F}^{2}[(p\cdot k')(k\cdot p')(-\frac{1}{2}+\sin^{2}\theta)^{2}+(p\cdot k)(k'\cdot p')\sin^{4}\theta)].
\end{equation}
To calculate the decay width, we will integrate over $p'$, $k$ and
$k'$, hence we can make use of the symmetry between $k$ and $k'$ to
write the squared amplitude as
\begin{equation}\label{amplitude}
\sum_{spin}\mathcal{M}\mathcal{M}^{*}=128G_{F}^{2}(p\cdot k')(k\cdot p')\left[\left(-\frac{1}{2}+\sin^{2}\theta_W\right)^{2}+\sin^{4}\theta_W\right].
\end{equation}
The decay width is formally given by
\begin{eqnarray}\label{Gamma1}
\Gamma&=&\frac{1}{2E_{p}}\int\frac{d^{3}\vec{p'}}{(2\pi)^{3}2E_{p'}}\int\frac{d^{3}\vec{k}}{(2\pi)^{3}2E_{k}}\int\frac{d^{3}\vec{k'}}{(2\pi)^{3}2E_{k'}}\frac{1}{2}|\mathcal{M}|^{2}(2\pi)^{4}\delta^{4}(p-p'-k'-k)\\
&=&\frac{8G_{F}^{2}}{(2\pi)^{5}E_{p}}\left(\frac{1}{4}-\sin^{2}\theta_W+2\sin^{4}\theta_W\right)\int\frac{d^{3}\vec{p'}}{E_{p'}}\int\frac{d^{3}\vec{k}}{E_{k}}\int\frac{d^{3}\vec{k'}}{2E_{k'}}(p\cdot k')(k\cdot p')\delta^{4}(p-p'-k'-k).\nonumber
\end{eqnarray}
Here $\theta_W$ is the Weinberg angle.

In the main part of this section, we will focus on the calculation
of integral by temporarily forgetting the overall coefficient.
Because we are interested in high-energy neutrino decay, the masses
of electron and positron will be neglected in our calculation. Then
the integral in \eqref{Gamma1} can be briefly rewritten as
\begin{eqnarray}
\Gamma_{\ab}&=&\int\frac{d^{3}\vec{p'}}{E_{p'}}\int\frac{d^{3}\vec{k}}{E_{k}}\int\frac{d^{3}\vec{k'}}{2E_{k'}}(p\cdot k')(k\cdot p')\delta^{4}(p-p'-k'-k)\nonumber\\
&=&\int\frac{d^{3}\vec{p'}}{E_{p'}}\int\frac{d^{3}\vec{k}}{E_{k}}\int d^{4}k'\delta\left(k'\cdot k'\right)|_{k'^{0}>0}(p\cdot k')(k\cdot p')\delta^{4}\left(p-p'-k'-k\right)\\
&=&\int\frac{d^{3}\vec{p'}}{E_{p'}}\int\frac{d^{3}\vec{k}}{E_{k}}\left[p\cdot(p-p'-k)\right]\left(k\cdot p'\right)\delta\left((p-p'-k)\cdot(p-p'-k)\right)|_{E_{p}-E_{p'}-k>0}.\nonumber
\end{eqnarray}

\begin{figure}
\begin{center}
\includegraphics[width=0.35\textwidth]{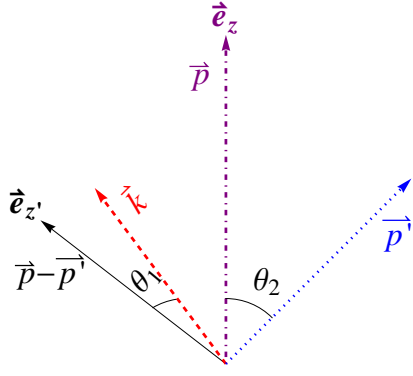}
\end{center}
\caption{(color online). The relative directions of vectors
$\vec{p}$, $\vec{p'}$, $\vec{p}-\vec{p'}$ and $\vec{k}$ in three
spatial dimensions. The zenith angle $\theta_{1}$ is defined as the
angle between $\vec{k}$ and $\vec{p}-\vec{p'}$, and $\theta_{2}$ as
the angle between $\vec{p}$ between $\vec{p'}$. As shown in the
picture, axes $\vec{\mathbf{e}}_{z}$ and $\vec{\mathbf{e}}_{z'}$
coincide with directions of $\vec{p}$ and $\vec{p}-\vec{p'}$
respectively. Projecting $\vec{k}$ on the plane perpendicular to
$\vec{\mathbf{e}}_{z'}$, we can define one azimuth angle
$\varphi_1$. Similarly, the other azimuth angle $\varphi_2$ can be
defined by projection of $\vec{p'}$ on the plane perpendicular to
$\vec{\mathbf{e}}_{z}$.}\label{fig-coord}
\end{figure}

Making use of relations $k\cdot k=k'\cdot k'=0$ and
\begin{equation}
k\cdot p'=k\cdot (p-k-k')=k\cdot p-\frac{(k+k')\cdot(k+k')}{2}=k\cdot p-\frac{(p-p')\cdot(p-p')}{2},
\end{equation}
we obtain
\begin{eqnarray}
\Gamma_{\ab}&=&\int\frac{d^{3}\vec{p'}}{E_{p'}}\int\frac{d^{3}\vec{k}}{k}\left[p\cdot(p-p')-p\cdot k\right]\left[k\cdot p-\frac{(p-p')\cdot(p-p')}{2}\right]\nonumber\\
&&\times\delta\left((p-p')\cdot(p-p')-2k\cdot(p-p')\right)\biggr|_{E_{p}-E_{p'}-k>0}\nonumber\\
&=&\int\frac{d^{3}\vec{p'}}{E_{p'}}\int kdk\sin\theta_{1}d\theta_{1} d\varphi_{1}\left[p\cdot(p-p')-E_{p}E_{k}+\vec{p}\cdot\vec{k}\right]\nonumber\\
&&\times\left[E_{p}E_{k}-\vec{p}\cdot\vec{k}-\frac{(p-p')\cdot(p-p')}{2}\right]\nonumber\\
&&\times\delta\left((p-p')\cdot(p-p')-2E_{k}(E_{p}-E_{p'})+2\vec{k}\cdot(\vec{p}-\vec{p'})\right)\biggr|_{E_{p}-E_{p'}-k>0}.
\end{eqnarray}
Here $\theta_{1}$ is defined as the angle between $\vec{k}$ and
$\vec{p}-\vec{p'}$. We will define $\theta_{2}$ as the angle between
$\vec{p}$ and $\vec{p'}$. The relative directions and angles between
the relevant momentum vectors are depicted in figure
\ref{fig-coord}.
It is convenient to express the inner product of $\vec{p}$ and
$\vec{k}$ in terms of the new coordinates,
\begin{equation}\label{pk}
\vec{p}\cdot\vec{k}=pk\frac{\cos\theta_{1}(p-p'\cos\theta_{2})+p'\sin\theta_{1}\cos\varphi_{1}\sin\theta_{2}}{|\vec{p}-\vec{p'}|}.
\end{equation}
In terms of the new coordinates, the integration takes the form
\begin{eqnarray}
\Gamma_{\ab}&=&\int\frac{d^{3}\vec{p'}}{E_{p'}}\int kdk\sin\theta_{1}d\theta_{1}d\varphi_{1}\biggl[p\cdot(p-p')-E_{p}E_{k}\nonumber\\
&&+pk\frac{\cos\theta_{1}(p-p'\cos\theta_{2})+p'\sin\theta_{1}\cos\varphi_{1}\sin\theta_{2}}{|\vec{p}-\vec{p'}|}\biggr]\nonumber\\
&&\times\left[E_{p}E_{k}-pk\frac{\cos\theta_{1}(p-p'\cos\theta_{2})+p'\sin\theta_{1}\cos\varphi_{1}\sin\theta_{2}}{|\vec{p}-\vec{p'}|}-\frac{(p-p')\cdot(p-p')}{2}\right]\nonumber\\
&&\times\delta\left((p-p')\cdot(p-p')-2E_{k}(E_{p}-E_{p'})+2k|\vec{p}-\vec{p'}|\cos\theta_{1}\right)\biggr|_{E_{p}-E_{p'}-k>0}\nonumber\\
&=&\int\frac{d^{3}\vec{p'}}{E_{p'}}\int kdkd\varphi_{1}\int^{1}_{-1}dx\biggl[p\cdot(p-p')-E_{p}E_{k}\nonumber\\
&&+pk\frac{x(p-p'\cos\theta_{2})+p'\sqrt{1-x^{2}}\cos\varphi_{1}\sin\theta_{2}}{|\vec{p}-\vec{p'}|}\biggr]\nonumber\\
&&\times\left[E_{p}E_{k}-pk\frac{x(p-p'\cos\theta_{2})+p'\sqrt{1-x^{2}}\cos\varphi_{1}\sin\theta_{2}}{|\vec{p}-\vec{p'}|}-\frac{(p-p')\cdot(p-p')}{2}\right]\nonumber\\
&&\times\delta\left((p-p')\cdot(p-p')-2E_{k}(E_{p}-E_{p'})+2k|\vec{p}-\vec{p'}|x\right)\biggr|_{E_{p}-E_{p'}-k>0}.
\end{eqnarray}

With the newly introduced variable $x=\cos\theta_{1}\in[-1,1]$, we
note that the function
\begin{equation}
f(x)=(p-p')\cdot(p-p')-2k(E_{p}-E_{p'})+2k|\vec{p}-\vec{p'}|x
\end{equation}
has a single root
\begin{equation}
x_{0}=\frac{2k(E_{p}-E_{p'})-(p-p')\cdot(p-p')}{2k|\vec{p}-\vec{p'}|}
\end{equation}
and its first-order derivative
\begin{equation}
f'(x)=2k|\vec{p}-\vec{p'}|.
\end{equation}
Therefore we can integrate $x$ out of the delta function and
quickly obtain
\begin{eqnarray}\label{Gamma2}
\Gamma_{\ab}&=&\int\frac{d^{3}\vec{p'}}{E_{p'}}\int kdkd\varphi_{1}\biggl[p\cdot(p-p')-kE_{p}\nonumber\\
&&+kp\frac{x_{0}(p-p'\cos\theta_{2})+p'\sqrt{1-x_{0}^{2}}\cos\varphi_{1}\sin\theta_{2}}{|\vec{p}-\vec{p'}|}\biggr]\nonumber\\
&&\times[E_{p}k-pk\frac{x_{0}(p-p'\cos\theta_{2})+p'\sqrt{1-x_{0}^{2}}\cos\varphi_{1}\sin\theta_{2}}{|\vec{p}-\vec{p'}|}-\frac{(p-p')\cdot(p-p')}{2}]\nonumber\\
&&\times\frac{1}{2k|\vec{p}-\vec{p'}|}\biggr|_{E_{p}-E_{p'}-k>0}\nonumber\\
&=&\int\frac{d^{3}\vec{p'}}{2E_{p'}|\vec{p}-\vec{p'}|}\int dk\biggl\{2\pi\left[p\cdot(p-p')-kE_{p}+p\frac{kx_{0}(p-p'\cos\theta_{2})}{|\vec{p}-\vec{p'}|}\right]\biggl[E_{p}k\nonumber\\
&&-p\frac{kx_{0}(p-p'\cos\theta_{2})}{|\vec{p}-\vec{p'}|}-\frac{(p-p')\cdot(p-p')}{2}\biggr]-\pi\frac{(pp')^{2}(k^2-k^2x_{0}^{2})\sin^{2}\theta_{2}}{|\vec{p}-\vec{p'}|^{2}}\biggr\}\biggr|_{E_{p}-E_{p'}-k>0}
\end{eqnarray}

One may check that the integrand of \eqref{Gamma2} is independent of
$\varphi_{2}$. Its dependence on $k$ is quite simple. Its domain of
integration is determined by $-1<x_{0}<1$, $E_{p}-E_{p'}-k>0$, or
namely
\begin{eqnarray}
\frac{(p-p')\cdot(p-p')}{2(E_{p}-E_{p'})+2|\vec{p}-\vec{p'}|}<k&<&\frac{(p-p')\cdot(p-p')}{2(E_{p}-E_{p'})-2|\vec{p}-\vec{p'}|},\nonumber\\
E_{p}-E_{p'}&>&|\vec{p}-\vec{p'}|.
\end{eqnarray}
So we can integrate over variables $k$ and $\varphi_{2}$
straightforwardly. The calculation is a little tedious, yielding
\begin{eqnarray}\label{Gamma4}
\Gamma&=&\frac{8G_{F}^{2}}{(2\pi)^{5}E_{p}}\left(\frac{1}{4}-\sin^{2}\theta_W+2\sin^{4}\theta_W\right)\pi^{2}\int\int\frac{p'^2d p'dy}{6E_{p'}}\Bigl[3E_{p}E_{p'}^{3}\nonumber\\
&&+(-6 E_{p}^2+2p^{2}-3pp'y)E_{p'}^{2}+(3E_{p}^{3}-3E_{p}p^{2}-3E_{p}p'^{2}+8E_{p}pp'y)E_{p'}\nonumber\\
&&+3pyp'^{3}+(2E_{p}^{2}-2p^{2}-4p^{2}y^{2})p'^{2}+(3p^{3}y-3pE_{p}^{2}y)p'\Bigr].
\end{eqnarray}
The domain of integration is
\begin{eqnarray}
p^2+p'^2-2pp'y&<&(E_{p}-E_{p'})^{2},\label{domain1}\\
-1<y&<&1.\label{domain2}
\end{eqnarray}
In the above, we employed notation $y=\cos\theta_2$ and turned on
the overall factor in front of the integral as given by
\eqref{Gamma1}.

The expression \eqref{Gamma4} for decay width is one of our main
results. In section \ref{sec:exam}, we will apply it to several
examples with explicit dispersion relations. Before proceed, let us
make some comments on \eqref{Gamma4}.

First, we would like to show that the decay width \eqref{Gamma4} is
positive-definite, as one should have anticipated from its
definition \eqref{Gamma1}. For this purpose, it is enough to focus
on the integrand inside the square brackets, which can be
transformed to
\begin{eqnarray}
[\cdots]&=&3(E_{p}E_{p'}-pp'y)\left[(E_{p}-E_{p'})^2-(p^{2}+p'^{2}-2pp'y)\right]+2(E_{p}p'-E_{p'}p)^2\nonumber\\
&&+2pp'(1-y)\left[2E_{p}E_{p'}-pp'(1+y)\right].
\end{eqnarray}
Taking account of inequalities \eqref{domain1} and \eqref{domain2},
this expression is nonnegative if $E_{p}E_{p'}\geq pp'$. This
condition is well-satisfied if $E_{p}-p\geq0$, $E_{p'}-p'\geq0$ at
lower energy and $dE_{p}/dp\geq1$, $dE_{p'}/dp'\geq1$ in the energy
region of superluminal neutrino.

Second, we note that \eqref{domain1} puts a lower limit of
integration $y>[p^{2}+p'^{2}-(E_{p}-E_{p'})^2]/(2pp')$. In most
situations, we have $E_{p}-E_{p'}\leq p+p'$, then this limit is more
stringent than $y>-1$, and hence the domain of integration is simply
$[p^{2}+p'^{2}-(E_{p}-E_{p'})^2]/(2pp')<y<1$, leading to the reduced
decay width
\begin{eqnarray}
\Gamma_{E_{p}-E_{p'}\leq p+p'}&=&\frac{8G_{F}^{2}}{(2\pi)^{5}E_{p}}\left(\frac{1}{4}-\sin^{2}\theta_W+2\sin^{4}\theta_W\right)\pi^{2}\int\frac{p'^2dp'}{6E_{p'}}\frac{1}{24pp'}\Bigl\{5E_{p}^6\nonumber\\
&&-3\left(15E_{p'}^2+5p^2-3p'^2\right)E_{p}^4+8E_{p'}\left(10E_{p'}^2-6p'^2+9pp'\right)E_{p}^3\nonumber\\
&&-3E_{p}^2\left[15E_{p'}^4-6\left(3p^2-8p'p+3p'^2\right)E_{p'}^2-(p-p')^2\left(5p^2+10p'p-3p'^2\right)\right]\nonumber\\
&&-24E_{p}E_{p'}p\left[(2p-3p')E_{p'}^2+3(p-p')^2p'\right]+5E_{p'}^6+3E_{p'}^4\left(3p^2-5p'^2\right)\nonumber\\
&&-3E_{p'}^2(p-p')^2\left(3p^2-10p'p-5p'^2\right)-5(p-p')^4\left(p^2+4p'p+p'^2\right)\Bigr\}.
\end{eqnarray}

\section{Examples}\label{sec:exam}

In sections above, we have derived the kinematical threshold and
``decay width'' of the bremsstrahlung-like process
\eqref{bremsstrahlung} for general dispersion relations of neutrino.
This was done under the assumptions made in section \ref{sec:assum}.
To check our main results \eqref{threshold} and \eqref{Gamma4}, we
will apply them to muon decay process in subsection
\ref{subsec:muon} and to Cohen-Glashow model in subsection
\ref{subsec:cg}. As further applications, we will use them to study
some other models in subsections \ref{subsec:lw}, \ref{subsec:step},
\ref{subsec:hl}.

\subsection{Muon decay}\label{subsec:muon}

Process \eqref{bremsstrahlung} can be regarded as a three-body decay
process by weak interaction. It is analogous to muon decay in
standard model. So the basic test of our result \eqref{Gamma4} is
comparison with
$\mu(p)\rightarrow\nu_{\mu}(p')+\bar{\nu}_{e}(k')+e(k)$ by
neglecting particle masses in the final states and replacing
neutrino with muon in the initial state. Setting $E_{p'}=p'$,
$E_{p}^2=p^2+m_{\mu}^2$, we work out \eqref{Gamma4} directly
\begin{equation}
\Gamma=\frac{G_{F}^{2}m_{\mu}^6}{192\pi^3E_{p}}\left(\frac{1}{4}-\sin^{2}\theta_W+2\sin^{4}\theta_W\right).
\end{equation}
It is different from the muon decay width by a factor
$(1/4-\sin^{2}\theta_W+2\sin^{4}\theta_W)$. This factor should be
replaced by $1$ if we incorporate the charged current. So our result
exactly passes the muon decay test.

\subsection{Cohen-Glashow model}\label{subsec:cg}

The second test is to recover the result in \cite{Cohen:2011hx}. For
this purpose, we set $E_{p}^2=p^2(1+\delta)$,
$E_{p'}^2=p'^2(1+\delta)$ in \eqref{Gamma4} and get
\begin{equation}\label{cg}
\Gamma=\frac{4}{7}\times\frac{G_{F}^{2}E_{p}^5\delta^3}{192\pi^3}\left(\frac{1}{4}-\sin^{2}\theta_W+2\sin^{4}\theta_W\right).
\end{equation}
The decay width of \cite{Cohen:2011hx} can be numerically
reproduced\footnote{This corrects the wrong claim in the first
version of our manuscript, because the modification of squared
amplitude changes the overall factor in the decay width.} by taking
$\sin^{2}\theta_W\simeq1/4$.

Another main result of this paper is the threshold
\eqref{threshold}. Applying this threshold condition to the model of
\cite{Cohen:2011hx}, we find
\begin{equation}
E_{\th}=\frac{2m_{e}+m_{\nu}}{\sqrt{1-\frac{1}{1+\delta}}}\simeq\frac{2m_{e}}{\sqrt{\delta}},
\end{equation}
the same as the threshold in \cite{Cohen:2011hx}.

\subsection{Mass-dependent Lorentz violation}\label{subsec:lw}

\begin{figure}
\begin{center}
\includegraphics[width=0.45\textwidth]{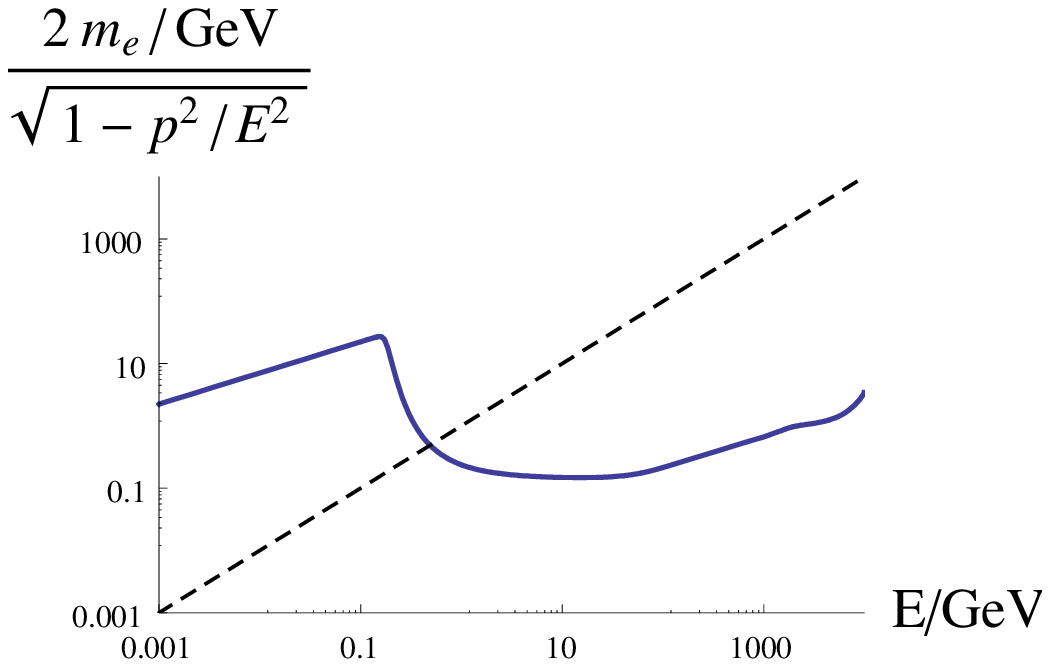}
\includegraphics[width=0.45\textwidth]{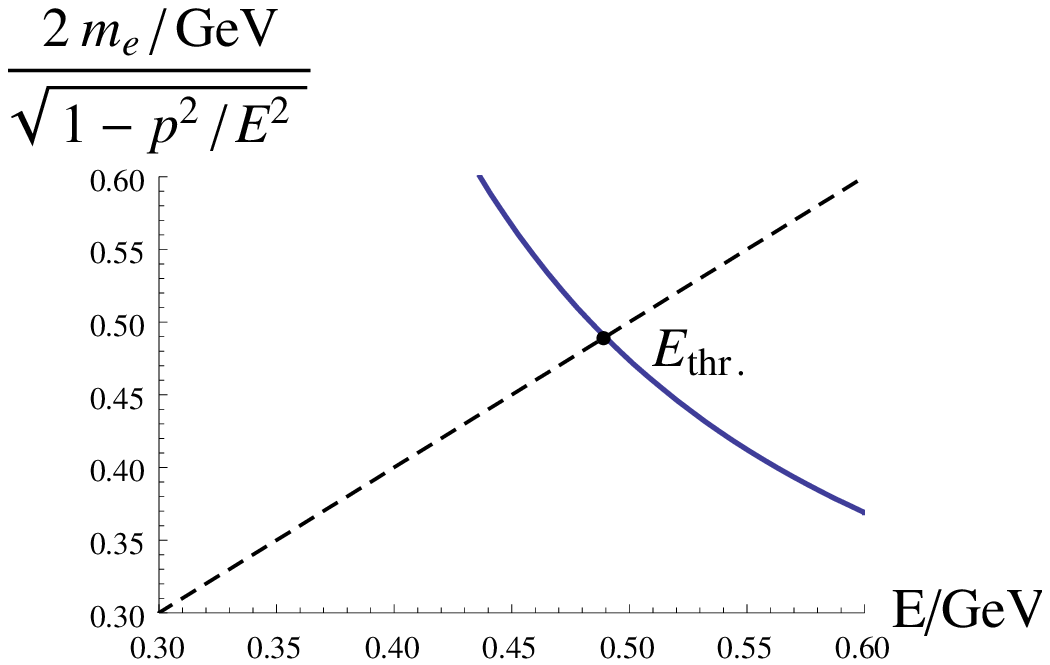}
\end{center}
\caption{(color online). Numerical solution of the threshold for a
toy model of mass-dependent Lorentz violation. The solid blue line
depicts function $2m_e(1-p^2/E_{p}^2)^{-1/2}$. It crosses the dashed
black line at the threshold energy, as highlighted by a black dot in
the right graph. The electron/positron mass is set to $0.5\MeV$. The
toy model and other parameter values are chosen to be the same as in
\cite{Li:2011ue}.}\label{fig:lw}
\end{figure}

In ref. \cite{Li:2011ue}, two of the authors proposed the
mass-dependent Lorentz violation scenario to explain the observed
neutrino velocity as a function of energy. In this scenario, the
mass-energy relation of neutrino has the form
\begin{equation}\label{m-E}
1-v^2=\lambda-f(\lambda),\quad\lambda=m^2/E^2
\end{equation}
where $f(\lambda)$ is a model-dependent function. The new function
$f(\lambda)$ is useful phenomenologically, because we can get its
information directly from experiments which constrain neutrino
velocity as a function of energy. With the definition of group
velocity $v=dE/dp$, relation \eqref{m-E} can be taken as a
differential equation and integrated into dispersion relation
\begin{equation}
p=\int_{m}^{E}\frac{d\tilde{E}}{\sqrt{1-\frac{m^2}{\tilde{E}^2}+f\left(\frac{m^2}{\tilde{E}^2}\right)}}.
\end{equation}
A concrete toy model of mass-dependent Lorentz violation was devised
in \cite{Li:2011ue}, well fitting observational data of neutrino
velocity. For the toy model and parameters given in
\cite{Li:2011ue}, we combine this relation with eq.
\eqref{threshold}, and numerically get the threshold energy at about
$0.5\GeV$. This is illustrated in figure \ref{fig:lw}. This
threshold is higher than that of the Cohen-Glashow model
\cite{Cohen:2011hx}.

Because the dispersion relation takes a very complicated form, it is
difficult to work out the decay width in this toy model, even
numerically. As an alternative, we will deal with a simplified model
in subsection \ref{subsec:step}.

\subsection{Velocity of step form}\label{subsec:step}

In the toy model of ref. \cite{Li:2011ue}, the dependence of
velocity on energy looks like a delta function, and well explains
the observational data of neutrino velocity. But it is very
difficult to work out the decay width \eqref{Gamma4} for that model.
As an alternative, let us study a model in which the neutrino's
velocity depends on energy in a step form
\begin{equation}\label{step}
dp/dE=1+\left(\frac{1}{\sqrt{1+\delta}}-1\right)H(E-E_{c}).
\end{equation}
Here $H(x)$ is a Heaviside unit step function and $E_{c}$ is an
critical energy.

\begin{figure}
\begin{center}
\includegraphics[width=0.45\textwidth]{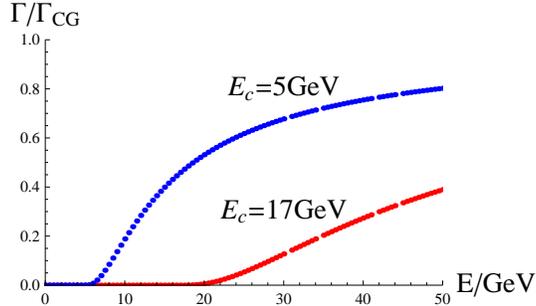}
\end{center}
\caption{(color online). The dependence of decay width
\eqref{Gamma4} on neutrino energy $E_{p}$ in model \eqref{step}. We
have fixed $\delta=2\times10^{-5}$, $E_{c}=5\GeV$ for the blue
points and $E_{c}=17\GeV$ for the red points.}\label{fig:step}
\end{figure}

Fixing $\delta=2\times10^{-5}$, we numerically computed the decay
width and got the results in figure \ref{fig:step}. From the figure
we can see the value of decay width becomes closer and closer to
\eqref{cg} as $E$ increases. The figure also tells us that the decay
width gets larger if neutrino is faster than light in a wider energy
range. Reversing the logic, the decay width can be suppressed by
narrowing the energy range in which the neutrino is faster than
light. Perhaps this is realizable torturously if the dependence of
neutrino velocity on energy takes a comb-like form.


\subsection{Horava-Lifshitz model}\label{subsec:hl}

Motivated by Horava-Lifshitz theories, ref. \cite{Mohanty:2011rm}
has studied the dispersion relation for neutrinos of the
form
\begin{equation}\label{hl0}
E^2=p^2+m^2+\eta'p^2+\frac{\eta p^4}{M^2}.
\end{equation}
Here the mass of neutrino $m$ is negligible. When the $\eta'$
correction dominates, this model reduces to the Cohen-Glashow model.

When the $\eta$ correction dominates, dispersion relation
\eqref{hl0} becomes
\begin{equation}\label{hl2}
E^2=p^2+\frac{\eta p^4}{M^2}
\end{equation}
with energy-dependent velocity. For this form of dispersion
relation, we can calculate the decay width with \eqref{Gamma4}. To
the leading order of $\eta$, it is\newpage
\begin{equation}\label{hl}
\Gamma=\frac{1665}{2002}\times\left(\frac{E_{p}^2}{M^2}\right)^3\times\frac{G_{F}^{2}E_{p}^5\eta^3}{192\pi^3}\left(\frac{1}{4}-\sin^{2}\theta_W+2\sin^{4}\theta_W\right).
\end{equation}
Remembering that the neutrino velocity is $v^2-1\simeq3\eta
p^2/M^2$, it is convenient to take the notation $\delta=3\eta
E_{p}^2/M^2$ and rewrite decay width \eqref{hl} as
\begin{equation}\label{hl}
\Gamma=\frac{185}{6006}\times\frac{G_{F}^{2}E_{p}^5\delta^3}{192\pi^3}\left(\frac{1}{4}-\sin^{2}\theta_W+2\sin^{4}\theta_W\right).
\end{equation}
At energies relevant to the OPERA experiment, such a decay width is
one or two orders smaller than that of the Cohen-Glashow model. This
ratio of suppression is consistent with the results of ref.
\cite{Mohanty:2011rm}. A naive application of \eqref{threshold} to
\eqref{hl2} yields $\eta p_{\th}^4/M^2=(2m_{e}+m_{\nu})^2$, which
gives the threshold energy in leading order
\begin{equation}
E_{\th}\simeq\frac{\sqrt{2m_{e}M}}{\eta^{1/4}}.
\end{equation}
It is unsafe to take $E_{\th}\simeq2m_{e}\sqrt{3/\delta}$, because
here $\delta$ is energy-dependent.

\section{Conclusion}\label{sec:con}

In this paper, under the assumptions enumerated in section
\ref{sec:assum}, we studied the kinematic threshold and decay width
of superluminal neutrinos for the bremsstrahlung-like process
\eqref{bremsstrahlung}. This was done for general dispersion
relations of neutrino, without resorting to any nontrivial frame
such as the effective ``rest frame''. The main results are
represented by eqs. \eqref{threshold} and \eqref{Gamma4}. Our
results confirmed and generalized the previous results in
\cite{Cohen:2011hx,Bi:2011nd}.

Before concluding this paper, we would like to make some relevant
remarks on the assumption of squared amplitude, which leaves a loose
end for future investigation. As has been emphasized in section
\ref{sec:assum}, when calculating the decay width, we assumed that
the squared amplitude is the same as that in standard model.
Strictly speaking, this is not always a consistent assumption when
general dispersion relations are involved. In general, dispersion
relations will enter into both kinematics and dynamics of particle
physics. However, without an assumption on the squared amplitude, we
cannot do any calculation about decay width. At the same time, since
the deviation of neutrino dispersion relation is not too far from
special relativity, we expect that the deformed amplitude should not
deviate significantly from the standard model. Therefore, our
assumption \eqref{amplitude} is not only necessary but also natural
to some extent. We thus expect our result provides a good estimation
of decay width in order of magnitude. Of course, further
investigation is required to confirm this expectation and improve
the present situation. See ref. \cite{Bezrukov:2011qn} for a recent
progress along this direction.

Another interesting project is employing our general results to rule
out more phenomenological models and hunt for viable models, as
shown by some examples in section \ref{sec:exam}. We feel this
project will be challenging but rewarding, given the importance of
special relativity in modern physics.


\acknowledgments{We are grateful to Xiao-Jun Bi, Tianjun Li and
Xiao-Dong Li for valuable discussion. T.W. would like to thank the
Kavli Institute for Theoretical Physics China for hospitality where
this work was done during a program on String Phenomenology and
Cosmology. M.L. is supported by the NSFC grants No.10535060,
No.10975172 and No.10821504, and by the 973 program grant
No.2007CB815401 of the Ministry of Science and Technology of China.
T.W. is supported by the NSFC grant No.11105053.}

\end{document}